\documentclass[a4paper,12pt]{article}
\usepackage[utf8]{inputenc}

\usepackage{xcolor}
\usepackage{amssymb}
\usepackage{lipsum}
\usepackage{amsmath}
\usepackage{bm}
\usepackage[authoryear]{natbib}  
\newcommand{\sign}{\text{sgn}}

\makeatletter

\title{Order flow and price formation}

\author{Fabrizio Lillo\\
		{\small University of Bologna and  Scuola Normale Superiore,}\\ Italy \\
	}

\begin{document}
\maketitle

\begin{abstract}
I present an overview of some recent advancements on the empirical analysis and theoretical modeling of the process of price formation in financial markets as the result of the arrival of orders in a limit order book exchange.  After discussing critically the possible modeling approaches and the observed stylized facts of order flow, I consider in detail  market impact and transaction cost of trades executed incrementally over an extended period of time, by comparing model predictions and recent extensive empirical results. I also discuss how the simultaneous presence of many algorithmic trading executions affects the quality and cost of trading.

\end{abstract}

\section{Introduction}

Understanding the price formation process in markets is of paramount importance both from an academic and from a practical perspective. Markets can be seen as a collective evaluation system where the `fair' price of an asset is found by the aggregation of information dispersed across a large number of investors. Non informed investor (roughly speaking, intermediaries and market makers) also participate to the process, in the attempt of profiting from temporal or `spatial' (i.e. across market venues or assets) local imbalance between supply and demand, thus acting as counterparts when liquidity is needed. 

Order submission and trading constitute the way aggregation of information is obtained. The process through which this information is impounded into price is highly complex and might depend on the specific structure of the investigated market. Prices emerge as the consequence of the arrival of orders, which in turn are affected, among other things, by the recent dynamics of prices. Despite the fact this feedback process is of paramount importance, the complexity of the process is only partial understood and many different models are able to provide only a partial description of it. 

The two main components of the price formation process are order flow and market impact. The former refers to the dynamical process describing the arrival of buy and sell orders to the market. As detailed below, this is in general a complicated process whose modelization is challenging because of the high dimensionality and the presence of strong temporal correlations. Market impact is, broadly speaking, the correlation between an incoming order and the subsequent price change. Since in each trade there is a buyer and a seller, it is not a priori obvious whether a given trade should move on average the price up or down. Considering the role of information on prices, one can advance few alternative explanations on the origin of market impact (for a more detailed discussion on this point, see \citep{BFL}): 
\begin{itemize}
\item {\bf Trades convey a signal about private information.} The arrival of new private information causes trades, which cause other agents to update their valuations, which changes prices. 
\item {\bf Agents successfully forecast short-term price movements and trade accordingly.} Thus there might be market impact even if these agents have absolutely no effect on prices. In the words of Hasbrouck `orders do not impact prices. It is more accurate to say that orders forecast prices'.
\item {\bf Random fluctuations in supply and demand.} Fluctuations in supply and demand can be completely unrelated to information, but the net effect regarding market impact is the same. In this sense impact is a completely mechanical (or statistical) phenomenon.
\end{itemize}

In the first two explanations, market impact is a friction but it is also the mechanism that let prices adjust to the arrival of new information. In the third explanation, instead, market impact is unrelated to information and  may merely be a self-fulfilling prophecy that would occur even when the fraction of informed traders is zero. Identifying the dominating mechanism  in real markets is therefore of fundamental importance to understand price formation.

Price formation and market impact are very relevant also from the practitioner perspective of minimizing transaction costs. For medium and large size investors, the main source of trading costs is the one associated with market impact, since by executing progressively an order in response to a given trading decision, the price is moved in a direction adverse to the trader and the later trades/orders become more and more expensive. Minimizing market impact cost by designing optimal execution strategies is an active field of research in academia and industry \citep{AC}. 

 Market impact is thus a critical quantity to understand the informativeness of a trade as well as the cost for the trader, but its nature and properties are still vigorously debated.  Also the empirical analysis and characterization of price formation and order flow is challenging, despite the availability of very high-resolution market data. This is due in part to the difficulty of controlling several potential confounding effects and biases and in part to the fact that market data are often not sufficient to answer some fundamental questions. However recent years have witnessed a booming increase in the number of empirical studies of market impact and transaction cost analysis of algorithmic executions, and we are now able to model them with a great level of accuracy and to dissect the problem under different conditioning settings.

In this subchapter I will review of some of these recent advancements in the modeling and empirical characterization of order flow, price formation, and market impact. I will focus on a specific, yet widespread, market mechanism namely the Limit Order Book, which is presented in section \ref{sec:lob}. 
In section \ref{sec:approaches} I will present an overview of the different modeling approaches to order flow and price formation, clarifying the different choices that the modeler has and why and when some should be preferred to others. Section \ref{sec:orderflow} reviews some results on order flow modeling and in Section \ref{sec:crossimpact} I will consider cross-impact, i.e. how the price of an asset responds to trades (and orders) executed on a different asset. The study of cross-impact is important when a portfolio of assets is liquidated, since cross-asset effects can deteriorate the quality of the trade if not properly included in the optimal execution scheme. 
Section \ref{sec:metaorder} presents empirical evidences and theoretical results on the market impact of metaorders, i.e. sequences of orders sent by the same trader as a consequence of a unique trading decision. 
Section \ref{sec:coimpact} discusses the problem of the simultaneous presence of many metaorders and how market impact behaves under aggregation. The response of price to multiple simultaneous metaorders has been recently termed co-impact and its characterization and modeling is important to study the effect of crowding on price dynamics and cost analysis.
Finally, in Section \ref{sec:conclusions} I will briefly present some open problems in the field of order flow and price formation and I will delineate few possible research avenues.

\section{The Limit Order Book}\label{sec:lob}

Market microstructure is, by definition, very specific about the actual mechanism implemented in the investigated market, because it can affect the price formation process. Financial markets are characterized by a variety of structures, and attempting to make a classification is outside the scope of this subchapter. In the following the focus will be on the most popular market mechanism, namely the Limit Order Book (LOB). A LOB, used actively also outside finance, is a mechanism for double auction and it is essentially a queuing system. Traders can decide to send their order (to buy or to sell) in two different ways: either they require to buy or sell a certain amount of shares at the best available price or they specify also the worst price at which they are willing to trade, thus the highest price for a buy or the lowest price for a sell. In the first case they send a {\it market order} (or, equivalently, a crossing limit order)
and, unless there is no one on the opposite case, the order is executed and leads to a {\it transaction}. In the second case they send a {\it limit order}, where the specified price is called the {\it limit price}, and, if no one is on the opposite side with the same (or more favorable) price, the limit order is stored in a queue of orders at the limit price. An agent can decide to cancel a limit order at any time, for example if the price moves in an adverse direction.  At any time, the highest standing limit price to buy (sell) is called bid (ask) or best bid (best ask). The mean price between the bid and the ask is the {\it midprice} and the difference is the {\it spread}. 
Orders arrive and are canceled asynchronously in the market and what is normally called `the price' is something in between the best ask and the best bid. However, from the above description it is clear that at a certain time there is not a unique price in the market.

Broadly speaking three modeling approaches have been pursued: (i) econometric models, fitting for example large dimensional linear models on market data (queues, prices, order arrivals); (ii) statistical models of the LOB, where orders arrive in the market as a random point process and the resulting properties of the price is studied; (iii) computational agent based models, where a set of heterogeneous agents trade in a realistic environment, such as a LOB. I will mostly focus on the first two approaches, despite the fact the third approach often provide important insights, especially for testing alternative policy measures.

\section{Modeling approaches}\label{sec:approaches}

Modeling order flow and price formation is a challenging task because of the complexity of the system and the large number of variables potentially involved. The modeler has different choices to make, which in part depend on the available data and methods, but more often depend on the objectives of the model.

The first choice is whether to work in continuous or in discrete time. The first option is the most complete, i.e. it does not discard any information of the process of price formation. Inter-event times can, in fact, provide relevant information on the event is going to occur. For example, the price change triggered by a trade can depend on the time elapsed from the last trade. The modeling in discrete time disregards this information but allows to use all the machinery of discrete time series analysis (ARMA, VAR, etc).  Discrete time modeling can be deployed either by advancing the clock by one unit any time a specific event occurs, for example a trade or an order arrival, or by considering a finite interval of physical time, say 1 second, and by considering aggregated quantities (e.g. average or end-of-period LOB, total order flow, one second price return, etc). 

Let us consider first the continuous time approach and let $K$ the number of available limit prices\footnote{Following \citep{contLOB}, we consider $K$ large enough that it is unlikely that in the considered period orders are placed outside the grid.}. Denoting with $p_t^i$ and $q_t^i$, ($i=1,...,K$) the price and the number of shares on the $i-$th limit price at time $t$, the LOB dynamics is described by the continuous-time process ${\mathcal L}_t=(p_t^i, q_t^i: i=1,...,K)$. The order flow is described by the multivariate marked point process whose components are the intensity of limit orders ($\lambda_t^i$), cancellations ($\nu_t^i$), and buy and sell market orders ($\mu_t^b$ and $\mu_t^s$). The marks correspond to the volumes of the order, but for expositional simplicity we will assume that all the orders have unitary volume. In general the rates are not constant in time but can depend on the past history of the order flow, on the state of the order book (${\mathcal L}_{\{s<t\}}$), and possibly on other covariates. Let us call ${\mathcal O}_t$ the multivariate point process generated by the intensities $(\lambda_t^i,\nu_t^i,\mu_t^b,\mu_t^s: i=1,...,K)$  and fully describing the order flow.  

It is important to stress that the state of the LOB at a given time is {\it completely determined} by the past order flow, plus some initial condition. In other words, once we choose an observable price $p_t$ as reference (for example the midprice, the microprice, the ask), there exists a deterministic function $F$ such that 
\begin{equation}
\Delta p_t\equiv p_t-p_{t-\tau}=F({\mathcal L}_{t-\tau}, {\mathcal O}_{s\in (t-\tau, t)})
\end{equation}
Thus, from a purely econometric point of view, one could simply model the point process process ${\mathcal O}_t$. This type of models is  often analytically tractable and, for this reason, it has been explored in the past twenty years in several papers. The Zero Intelligence (or Santa Fe) model of \citep{daniels2003quantitative} and the model in \citep{contLOB}, for example, consider independent Poisson processes for the different components of ${\mathcal O}_t$. In order to include memory of the past order flow, \citep{Abergel} considers instead a multivariate Hawkes processes able to describe auto- and lagged cross-correlation between the different components of the order flow.

The observable reference price in the LOB might not the fully reflect the economic conditions of the firm. For this reason, many models postulate the existence of an unobservable {\it efficient} price, which typically follows a semimartingale dynamics. Market data (e.g. trade or mid price) are a noisy version of the efficient price and a lot of econometric effort is devoted to remove the microstructure noise either to filter it or to estimate from ultra high frequency data some of its statistical properties (for example the volatility) useful in applications such as option pricing or risk management.  

Although order flow determines uniquely (observable) price changes, it is possible that a better model (in terms, for example, of explained variance) is obtained by considering the order flow intensities as dependent on LOB state ${\mathcal L}_t$ or of a function of it, such as the reference price $p_t$. The reason is that, in general, the relation between intensities and past order flow is strongly non-linear and high dimensional. On the contrary, simpler and easier to estimate parametric models can be chosen by identifying the drivers that supposedly influence real traders decision to submit a specific type of order\footnote{A recent alternative approach is to use modeling approach suited for high-dimensional non linear models, such as Deep Neural Networks. Even in these cases however it might be better to use LOB state rather than past order flow  to forecast the LOB state at a future time. For example, Deep Learning has been used to forecast short term price movement from LOB state and recent order flow (see, for example, \citep{Sirignano}).}. For example, real traders likely decide when and where to place an order taking into account the LOB state and the price. Thus one could use a model, which instead of modeling autonomously the order flow,  makes the intensities dependent on the state of the LOB or of part of it \citep{Rosenbaum}. 

Choosing to build models using functions of the order flow could be also useful when deciding to restrict the dimensionality of the problem and restricting it to a subpart of the order flow (and of LOB). The reasons for this choice are manyfold: either for data availability (especially in the old times), for purely statistical reasons (dimensionality reduction and improved estimation), because one believes that some parts of the order flow (e.g. trades) might be more informative on price dynamics, or because we are interested in modeling a part of the order flow and its effect on price (for example our order flow in a real trading problem). In these cases the reduced model giving the price as a function of the (sub)order flow becomes stochastic and the randomness describes the effect of the unmodeled part of order flow. Following this line, one can take two approaches:

 1)  Treat the order flow as exogenous to the price. In this case, the model connects the considered part of the order flow to the price, but neglects the reverse effect, i.e. how price dynamics can affect order flow. Classical market microstructure models following this approach are the Roll model (and its generalization) and the Madhavan-Richardson-Roomans  \citep{MRR} model. More recently,  the Transient Impact Model (TIM, see \citep{Bouchaud,BFL}) and its generalizations with multiple propagators have been proposed to describe the relation between order flow and price. In a nutshell, the general TIM can be written in discrete time (see below for the continuous time version) as
\begin{equation}\label{eq:tim}
 p_t=\sum_{s<t} G_{\pi_{s}}(t-s) f(v_{s})+\xi_t + p_{-\infty}
\end{equation}
 where $v_{s}$ is the signed volume of the order at time $s$, $f(x)=\sign(x) h(|x|)$ with $h(.)$ a concave function\footnote{To avoid dealing with volume fluctuations, strongly dependent on LOB state, often it is chosen $f(x)=\sign(x)$. $\sign(x)$ denotes the sign function.}, as observed empirically \citep{lillo2003econophysics}, $\pi_s$ indicates the type of event at time $s$ (e.g. market order, limit order at a given price, etc), $G_{\pi_s}(t-s)$ is a function, termed {\it kernel} or {\it propagator}, quantifying the lagged effect of the event $\pi_s$ at time $s$ on the price at time $t$, and $\xi_t$ is a noise term describing the effect on price of all the orders which are not considered in the model. If the functions  $G_{\pi_s}$ are not constant, Eq. (\ref{eq:tim}) describes the {\it transient} nature of impact of event $\pi_s$, i.e. the fact that the effect of an order on price is not permanent, but declines with time. Many empirical analyses show that $G_{\pi_s}$ are slowly decaying functions, typically well fitted asymptotically by a power law function. The transient nature of impact can be related to the very persistent autocorrelations of order flow (see next section) and to the diffusivity and efficiency of prices (see \citep{BFL} for an extensive discussion). While the original TIM was considering only one type of events, namely market orders, subsequent works have included also limit orders and cancellations, while others have discriminated more finely between orders changing and not changing the price, since the (lagged) effect on price is shown to be different in these cases \citep{eisler,Taranto}.

The TIM describes trades that impact prices, but with a time dependent, decaying impact function $G(t)$. One can interpret the same model slightly differently. Considering  the model with one propagator associated with trades and taking $f(v_t)= \sign (v_t)\equiv \epsilon_t$, one can rewrite the model as
\begin{eqnarray}
\Delta p_t = G(1) (\epsilon_t-\hat \epsilon_t) +\tilde \xi_t ,\\
\hat \epsilon_t=-\sum_{s>0}\frac{G(s+1)-G(s)}{G(1)}\epsilon_{t-s}
\end{eqnarray}
with $\tilde \xi_t=\Delta \xi_t$. The quantity $\hat \epsilon_t$ can be seen as the (linear) predictor of trade sign given the past history of the signs and the model tells us that the deviation of the realized sign $\epsilon_t$ from an expected level $\hat \epsilon_t $ impacts the price linearly and permanently. If $\hat \epsilon_t$ is the best possible predictor of $\epsilon_t$, then the above equation leads by construction to an exact martingale for the price process. This model has been termed History Dependent Impact Model (HDIM) \citep{lilloFarmer2004,BFL,Taranto} and in the simple setting above is mathematically equivalent to the TIM when the best predictor is linear in the past order signs. \citep{Taranto} shows that as soon as one attempts to generalize the model to multiple event types, TIM and HDIM become no longer equivalent. In fact, the HDIM with different events can be rewritten as 
\begin{equation}
\Delta p_t=G_{\pi_t}(1)\left[\epsilon_t+\sum_{s<t}\frac{\kappa_{\pi_s,\pi_t}(t-s)}{G_{\pi_t}(1)}\epsilon_s\right]+\tilde \xi_t
\end{equation}
where $\kappa_{\pi_s,\pi_t}(t-s)$ is an influence kernel that depends on both the past event type $\pi_s$ and the current event $\pi_t$. The matrix of two point kernels makes the model more complicated to estimate (see \citep{Taranto}) while clearly HDIM reduces to TIM when $\kappa_{\pi_s,\pi_t}(t-s)$ is a function only of the triggering event $\pi_s$.

The approach taking as exogenous the order flow includes also other models connecting the order flow in a given time interval with the simultaneous price change. A paradigmatic example is given by \citep{ContKukanov14} which introduces a stylized model of the order book predicting a {\it contemporaneous} linear relation between the price change in a given time interval and a linear combination of level I order flow components (the Order Flow Imbalance or OFI).  The goodness of the model (for large tick stocks) is testified by the high $R^2$ empirically obtained in the linear regression between $\Delta p_t$ and OFI. 
 
2) The above approach, however, leaves the order flow as completely exogenous. The limits of this approach are evident for example when considering the negative lag response, i.e. the lagged cross correlation between past price returns and future order flow, i.e.
\begin{equation}\label{eq:response}
R(\tau)={\mathbb E}[f(v_t)(p_{t+\tau}-p_t)]
\end{equation}
with $\tau<0$. When the simple TIM model with only one propagator for trades is calibrated using $R(\tau)$ (or other equivalent methods) with $\tau>0$, it is observed that the predicted negative lag response is smaller than the empirical one, indicating, as intuitive, that a declining (increasing) price attracts in the future more buy (sell) trades \citep{Taranto}.  
 To overcome this problem, one jointly models the dynamics of price and order flow. The seminal work in this context is  \citep{Hasbrouck} who proposed a discrete time structural VAR model for the vector ${\bm x}_t=(\Delta p_t,f(v_t))'$, $v_t$ being the volume of the market order at time $t$, of the form 
\begin{equation}
 A_0 {\bm x}_t=\sum_{i=1}^p A_i {\bm x}_{t-i} +{\bm \xi}_t
\end{equation}     
and $A_0= \left(\begin{array}{cc}1& g \\0 & 1\end{array}\right)$ and $A_i$ are other $2\times 2$ matrices to be estimated. The parameter $g$ describes the immediate impact of a trade on price. 

This model has been generalized in several directions. First, instead of considering only market orders and a single reference price (Hasbrouck used the midprice),  \citep{Hautsch} considers a vector containing bid and ask prices, the queue volume at the first three quotes on either sides of the LOB, and two dummy variables indicating the occurrence of buy and sell trades. They modeled this ten dimensional vector ${\bm y}_t$ with the Vector Error Correction Model 
\begin{equation}
\Delta {\bm y}_t={\bm \mu}+\alpha \beta' {\bm y}_{t-1}+\sum_{i=1}^p \Gamma_i \Delta {\bm y}_{t-i}+{\bm \xi}_t
\end{equation}
where ${\bm \mu}$ is a constant vector, $\alpha$ and $\beta$ denote the loading and cointegrating matrices and $ \Gamma_i$ are parameter matrices.
Using impulse response function, they measured impact, separately on bid and ask prices, of the arrival of a limit order on a queue (for an approach based on TIM, see \citep{eisler}), as well of course the impact of the arrival of a market order. 
 
 The second generalization considered instead continuous time models. For example,  \citep{muzy} introduces an Hawkes process for the four dimensional counting process ${\bm P}_t=(T^+_t,T^+_t,N^+_t,N^-_t)'$, where the first two components describe the arrival of buy and sell market orders and the last two the upward/downward movements of the reference price. The model for the intensity ${\bm \lambda}_t$ reads
 \begin{equation}
 {\bm \lambda}_t= {\bm \mu}+\int_{-\infty}^t \Phi(t-s) d{\bm P}_s
 \end{equation}
 where ${\bm \mu}$ is a constant baseline intensity, $\Phi(t-s)$ is a $4\times 4$ matrix of kernels describing the lead-lag effects. Some care must be taken to model the immediate impact (the $g$ term in the Hasbrouck model above) by introducing a Dirac delta component in some elements of $\Phi(t-s)$.  While this model can be directly put in connection with the Hasbrouck's VAR, its generalizations can easily include other components of the order flow, such as limit orders and cancellations, or even to take into account order volume (see, for example, \citep{Rambaldi} discussed below). Moreover the continuous time approach allows to consider in the modeling the time between events, which are clearly neglected in discrete time approach \'a la Hasbrouck.

Understanding the relation between order flow and price is important for many reasons, such as to create realistic LOB simulators, to study the stability of markets under different rules, etc. However, it is often very relevant to study how price reacts to a specific sequence of orders generated by a specific trading decision, i.e. what we called a metaorder,  because this is related to transaction cost (mainly due to market impact) and to the release of private information into the prices. It is evident that, since we are neglecting a very large fraction of orders, those due to all other traders, the relation between price dynamics and order flow of a single metaorder will become very noisy and large samples are required to obtain clean measures. Section \ref{sec:metaorder} presents some empirical evidences on the market impact of metaorders and the price dynamics during their execution.

\section{Order flow} \label{sec:orderflow}

Order flow is the process describing the arrival of orders in the market. If this works with a LOB, then the order flow is the multivariate point process describing the arrival of market orders, limit orders, and cancellations. Since limit orders (and cancellations) are also characterized by a limit price, a component of the multivariate process should be associated with each limit price, making immediately the problem high-dimensional.

Different point process models of order flow have been proposed, ranging from (compound) Poisson processes \citep{daniels2003quantitative} to self exciting Hawkes processes \citep{Abergel,Rambaldi}.  Here we first review some of the empirical evidences and stylized facts observed in order flow which make challenging the development of a realistic model of the LOB. 

The first empirical evidence (even chronologically) is the so called {\it diagonal effect} \citep{biais} i.e. orders of a specific type are more likely to be observed just after orders of the same type. Interestingly, \citep{Rambaldi} extends this analysis by using a Hawkes process approach and shows that, by including volume into the analysis, the diagonal effect is markedly stronger for same-type same-size orders (see below).

The diagonal effect is the manifestation of a more significant regularity observed in real LOBs: components of the order flow are extremely persistent, i.e. long range autocorrelated in time. To present a specific example, consider only market orders, where volume is neglected and time is discretized in such a way that it increases by one unit each time a new market order arrives. Denoting with $\epsilon_t$ the sign of the $t$-th market order, being equal to $+1$ ($-1$) for a buy (sell) order, it has been empirically shown \citep{lilloFarmer2004,Bouchaud} that its autocorrelation function behaves asymptotically as $C(\tau)\equiv Cor[\epsilon_t,\epsilon_{t+\tau}]\sim \tau^{-\gamma}$ with $\gamma \in [0,1]$.
The empirical value of $\gamma\simeq 0.5$ shows that the market order sign is a long memory process, i.e. it lacks a typical time scale, with Hurst exponent $H=1-\gamma/2\simeq 0.75$.  A similar behavior has been observed for the other components of the order flow.

Several explanations have been proposed for this stylized fact, empirically observed in many different markets, asset classes, and time periods. The theories can be clustered in two classes: the first states that this is the effect of {\it herding}, i.e. several investors share the same view on the asset around the same time and trade accordingly. The second explanation is instead related to the fact that each trader creates an autocorrelated order flow and this is due to the practice of {\it order splitting}. Despite the fact that it is possible to create agent based models with either of the two mechanisms reproducing a correlated order flow, the assessment of the mechanism mainly responsible for this observation should be based on empirical evidences. \citep{tothJEDC}  proposes a method to disentangle the herding and splitting contributions to the autocorrelation. The idea to use labeled data, i.e. data where the identity of the trader sending the order is known (even if anonymized). 
 The autocorrelation function of order flow can be exactly decomposed as $C(\tau)=C_{split}(\tau)+C_{herd}(\tau)$
where the first (second) term is the contribution to the correlation considering only cases when the two market orders at time $t$ and $t+\tau$ were placed by the same (different) trader(s). To measure the relative importance of the two components, \citep{tothJEDC} uses brokerage data. Some exchanges provide  data where each order contains the coded identity of the broker who sent the order.  An extensive investigation of LSE data shows unambiguously that $C_{split}(\tau)$ explains always more than 75\% of $C(\tau)$ and, except for very short $\tau$ (one or two trades) the value is above 85\%. This empirical finding strongly indicates that order splitting is the main driver of the correlated order flow. Similar results are obtained when using data with agents rather than brokers. 

Market orders describe only part of the order flow dynamics. Among the several approaches to describe the full correlation structure of order flow (and price) we mention here the one using Hawkes processes. Generalizing a pioneering paper \citep{large},  \citep{jaisson} models level-I order book data by using 8-dimensional Hawkes process whose components are market, limit, cancel order (buy and sell), and mid-price changes (up and down). Using a non-parametric estimation method, their main finding is that the dominating driver of the process is self-excitation (i.e., once more, the diagonal effect). The only exceptions are the mid-price components for which cross-excitation effects are strongly dominating. Moreover there is a significant mean reversion of price, since present price changes trigger  price changes in the opposite direction. Interestingly most of the estimated diagonal kernels of the Hawkes process are slowly decreasing and well described by a power-law behavior,  consistent with the long memory described above.

This type of approach can be generalized in several directions. For example, \citep{munitoke} considers a full order book modeling using Hawkes process to disentangle the role and interaction between liquidity takers and providers. 
Another generalization considers the fact that orders are also characterized by a volume. Mathematically one can treat volumes as marks of the multivariate point process. Alternatively, when only few levels are considered in the analysis, one can bin the volume in $D$ groups and consider the volume process as the superposition of
$D$ unmarked point processes, each of which corresponds to
one of the possible $D$ values  that volume can take \citep{Rambaldi}. It is found that order size does matter, since kernels for different volume bins are quite different. Moreover large orders trigger cascade of small orders and small limit orders and cancellations strongly cross-excite, indicating hectic order re-positioning by market makers.

Despite the fact one can decide to model the (marked) multivariate process autonomously, obtaining as a `byproduct' the state of the LOB and, as a consequence, the price dynamics, it is likely that LOB state (but also recent price dynamics) describes better, or in a more parsimonious way, the local intensities of the orders arrival. The intuition is also related to the fact that traders often condition the decision of submitting an order to the state of the LOB. This suggests a class of models where intensities are a function of the LOB state. This approach has been pioneered in \citep{Rosenbaum} where orders arrivals are modeled as Poisson processes whose intensity is a function of the current state of the LOB. Thus the empirically observed autocorrelation of the order flow is seen as  a `consequence'  of the persistence of the queue size, but, conditionally on them, the arrival of orders follows a Poisson process. A natural extension of this model considers the order arrival intensity as a function both of the LOB state and of the past order flow. By using an Hawkes model with a kernel depending on both these variables, the State Dependent or Queue Reactive Hawkes models \citep{Pakkanen,Wu} have been proposed.

\section{Cross impact}\label{sec:crossimpact}

Up to now we have considered the market impact of trades and orders from a single asset. However, institutional investors rebalancing their portfolio very often trade simultaneously many assets.  Both the optimal execution problem and the assessment of transaction costs of metaorders should therefore take into account possible interactions between assets. 

Generically, there are  three sources of interaction: (i) statistical dependence  in asset prices, i.e. the well-known fact that returns of different assets are correlated; (ii) commonality in liquidity across assets \citep{Chordia}, i.e. the fact that, for example, the arrival rate of signed market (or limit) orders is correlated across assets, and (iii) quote revision effects, i.e. a trade in an asset can lead market makers to modify the bid and ask price in another related asset. The (lagged) correlation between price and order flow is termed {\it cross-impact}. 
As in the single asset case, the entire order flow completely determines the (reference) prices of the assets, thus one can trivially explain cross-impact (as well as self-impact) as a mere consequence of order flow dynamics and correlations.  However, when conditioning to a subset of the order flow (for example a market order or the child orders of a metaorder), or when the {\it future} price evolution is of interest, the dynamics becomes stochastic, because of the unmodeled part of the order flow and suitably modeling cross-impact becomes critical for predictions or ex-ante cost estimation. Under this conditioning, cross-impact can be dissected as the result of the three sources described above \citep{benzacross}.

Cross-impact has been empirically studied recently, see e.g. \citep{benzacross,schneider2018cross} and its role in optimal execution has been highlighted in \citep{mastromatteo2017trading,Gerry}. We review here some results obtained when considering the market order flow. First, there is a measurable cross asset effect between order flow and price as can be measured by the cross response function, which generalizes Eq. \ref{eq:response} as
\begin{equation}
R^{ij}(\tau)={\mathbb E}[f(v^i_t)(p^j_{t+\tau}-p^j_t)]
\end{equation}
between an order on asset $i$ at time $t$ and the price change of asset $j$ in $[t,t+\tau]$. $R^{ij}(\tau)$ is found to be different from zero and smaller than $R^{ii}(\tau)$ by a factor $\sim 5$ \citep{benzacross,schneider2018cross}.  To investigate the source of this lagged correlation \citep{schneider2018cross}, by investigating empirically the high frequency dynamics of Italian sovereign bonds traded in an double auction market, find evidence that  both lagged correlations of orders across assets and quote revisions play a role in forming cross-impact.  This result is obtaned by investigating the effect on price of bonds of isolated trades, i.e. trades on a bond such that no other trade is observed in other bonds a time window around it. This results indicates that both commonality in liquidity taking and price revision across assets are responsible for cross impact effects.

 The TIM can be easily extended to the multi asset case. Considering the continuous time version of the TIM,  the price of asset $i$ at time $t$ is
\begin{equation}\label{eq:multiTIM}
 p_t^i=p_0^i+\sum_j \int_0^t f^{ij}(\dot x_s^j) G^{ij}(t-s)ds+\int_0^t \sigma^i_s dW_s^i
\end{equation}
 where $f^{ij}(\dot x_s^j)$ is the (instantaneous) impact on the price of asset $i$ of trading asset $j$ at a rate $\dot x_s^j$, $G^{ij}(\cdot)$ is the decay kernel describing the lagged effect of trading on price, $\sigma^i_s$ is the volatility of asset $i$ and $W_s^i$ is a Wiener process. This model can be estimated on real data and it is found that: (i) $f^{ij}$ is non-linear and well described by a power law function with an exponent smaller than $1$ as for $f^{ii}$; (ii) the kernels $G^{ij}$ also display a power law behavior similar to $G^{ii}$, but with a significantly smaller amplitude; (iii)  the matrix $\{G^{ij}\}_{i,j=1,N}$ has a strong sectorial structure, similar to the one observed for returns \citep{benzacross,schneider2018cross}.  These regularities and the modeling can be successfully used to design optimal portfolio executions \citep{mastromatteo2017trading}.
 
Another important question is whether a model like (\ref{eq:multiTIM}) is always well posed or if there are trading strategies $\Pi=\{\bm{x}_t\}_{t \in [0.T]}$ allowing for price manipulation. More precisely, we remind that a \textit{round-trip trade} is a sequence of trades whose sum is zero, i.e. a trading strategy $\Pi$ with $\int_0^T{\dot{\bm{x}}_t \mathrm{d}t} = \bm{0}$. A \textit{price manipulation} is a round-trip trade $\Pi$ whose expected cost $C(\Pi)$ is negative and the principle of no-dynamic-arbitrage states that such a price manipulation is impossible. For the multi asset TIM this implies that 
\begin{equation}
C(\Pi) = \sum_{i,j}{ \int_0^T{ \dot{x}^i_t \mathrm{d}t \int_0^t{f^{ij}(\dot{x}^j_s) G^{ij}(t-s) \mathrm{d}s   }  } } \geq 0
\label{eq:nodynamicarb_ND}
\end{equation}
\citep{schneider2018cross} proves a series of theorems constraining the form of $f$ and $G$ in order to avoid price manipulation. In particular authors showed that for bounded decay kernels instantaneous cross-impact $f$ must be an odd and linear function of trading intensity and cross-impact from asset $i$ to asset $j$ must be equal to the one from $j$ to $i$. When a non vanishing bid-ask spread is considered, some inequalities between spread, maximum trading speed, and cross-impact asymmetry must be verified to avoid price manipulation.

\section{Market impact of metaorders}\label{sec:metaorder}


While the above described models generically describe the relation between order flow and price, it is often of practical and academic interest to study the price dynamics when  conditioning this relation to the execution of a (large) order by a specific trader following a single trading decision (a metaorder). The seminal paper of \citep{kyle1985continuous} shows that for a trader with insider information it is optimal to split the volume to be executed in many transactions to be executed incrementally over an extended period of time.

A part from the practical problem of minimizing transaction costs,  the relation between metaorder execution and price dynamics is relevant to understand how information is incorporated into price. In fact, a metaorder by definition corresponds to a trading decision, which in general is the response of the trader to a piece of information. For this reason, it is important to understand, not only how the price changes during the execution of the metaorder, but also the long term  level reached by the price when the transient effects due to the imbalance between supply and demand are dissipated. 

Note again the difference between the relation between order flow and price dynamics when one considers all market participants  or only the order flow generated by the trading decision of a single trader. As said above, price dynamics is a {\it deterministic} function of the order flow, 
while, when conditioning on the order flow of a specific trader, we expect a very noisy relation between signed volume and price change. However the objective here is not to have an high $R^2$ between them, but to answer the question: how much my trading activity consequent to a trading decision is going to affect {\it on average} the price?

 Measuring market impact of metaorders is typically quite complicated because it requires suitable data that cannot be inferred from public (e.g. market) data. In fact, it is necessary to have access to data where one can track the activity of a single trader (broker or investor) following a given trading decision. For this reason, most of the empirical researches on this topic has been performed by using trading data from a given institution or trading desk \citep{Torre,almgren2005direct,toth2011anomalous}. A part from the difficulty of accessing such data, this type of analyses runs the risk of being biased, since the sample is limited to a specific fund, which might have an idiosyncratic trading style. Market wide investigations of market impact of metaorders have been conducted by following two approaches. First, some exchanges exceptionally provide data where the coded identity of the market member is disclosed; thus by using suitable statistical methods, one can infer metaorders as sequences of trades/orders by the same member on the same asset with the same sign (see for example, \citep{moro2009market,Vaglica2010,tothlillo10}). The other approach requires the access to databases collected by specialized institutions and containing information about the metaorder executions of a large set of investors. The most important example is probably the dataset provided by ANcerno Ltd, a transaction cost analysis firm for institutional investors. According to some estimates, it accounts for more than 10\% of CRSP volume in US markets, thus providing a wide coverage of metaorder trading activity from many different institutional investors.

Methodologically there are two main problems in measuring market impact of metaorders. First, impact might depend on several conditioning variables, such as the market conditions at the time of the trade, the execution algorithm, etc., thus different conclusions might be drawn depending on the choice made. Second, market impact of metaorders is typically very noisy (see above), and, as a consequence, large datasets are required to obtain small error bars on the estimated impact. It is important to stress that market impact contributes as a {\it drift} term to the unperturbed dynamics of price. For this reason, in order to measure market impact it is fundamental to take into account the sign of the trade of the metaorder. 

The main quantity of interest is the {\it metaorder impact} defined as 
\begin{equation}\label{eq:impdef}
{\cal I}(Q,T)\equiv  {\mathbb E}[\epsilon \Delta \log p| Q,T] \end{equation}
where $\Delta \log p$ is the logprice change between the end and the start of the metaorder,  $Q$ is the size of the metaorder (in shares), $T$ is the metaorder duration (in seconds or in volume time, to minimize possible intraday effects), and $\epsilon$ is the sign of the metaorder (i.e $\epsilon=+1$ for a buy and $\epsilon=-1$ for a sell order).  Notice that ${\cal I}(Q,T)$ is directly related to the average impact cost of a metaorder execution. In fact, for an execution described by $\Pi=\{x_t\}_{t \in [0,T]}$, where $x_t$ is the asset position at time $t$, the expected implementation shortfall cost, i.e. the difference between the expected cost and the theoretical cost obtained by marking to market the trade with the initial price, is 
$C(\Pi)=\int_0^T \dot x_t  {\cal I}(x_t,t) ~dt$, 
where $\dot x_t$ is the time derivative of $x_t$ (i.e. the trading speed). Market impact is better described in terms of normalized quantities which also allows to consider different assets and different time periods in the same analysis. The first key quantity is the daily (or volume) fraction, defined as $\phi=Q/V$, where $V$  is the average daily traded volume\footnote{$V$ and $\sigma$ (see below) are typically estimated over the past $10\div25$ trading days, excluding the day when the metaorder is executed.}. The second quantity is the participation rate $\eta$, i.e the ratio between $Q$ and the volume traded in the market during the execution. The third one is the metaorder duration $T$, which can be obtained from $T=\phi/\eta$.

Remarkably, many empirical studies \citep{Torre,moro2009market,Zarinelli,toth2011anomalous,bershova2013non, waelbroeck2013market} seem to agree on the validity of the `square-root impact law', obtained when conditioning on the volume fraction of the metaorder
\begin{equation}\label{eq:squareroot}
{\cal I}(Q,T)\approx Y\sigma\sqrt{\phi}
\end{equation}
where $Y\simeq1$ is a numerical constant and $\sigma$ is the daily volatility of the asset. Eq. \ref{eq:squareroot} has been empirically shown also for disparate asset classes as options \citep{tothoption} and Bitcoin \citep{donierbitcoin}. This empirical relation is at first sight surprising: it indicates that the style of trading (for example using limit orders or market orders), the duration $T$ of the execution, the trading speed (i.e. the number of shares traded per unit time), etc, are not relevant!  These observations indicate that there must be some limitations to the validity of this `law'.  For example, the prefactor $Y$ might depend on the trading algorithm. 
 
 More recent and extensive empirical analyses \citep{Zarinelli} clarify the limits of the square-root impact law and highlight 
  some deviations. Specifically: 
\begin{itemize}
\item Considering a power law dependence on $T$ and $\eta$, \citep{Zarinelli} investigates the regression
\begin{equation}\label{eq:fitimpact}
{\cal I}(Q,T) = A~T^{\delta_T} \eta^{\delta_\eta} \cdot noise
\end{equation}
to measure the dependence of metaorder impact {\it separately} on participation rate and duration. The fitted exponents are $\delta_T=0.54\pm0.01$ and $\delta_\eta=0.52\pm 0.01$, and $A=0.207\pm 0.005$. The fact that both exponents are very close to $1/2$ indicates that ${\cal I}(Q,T)\approx \sqrt{\phi}$, at least as a first approximation, even when considering the effect of participation rate and duration.
\item By considering ${\cal I}(Q,T)$ as a function only of $\phi=Q/V$, it is clear that a logarithmic function fits the data better than a power law function; this indicates a linear behavior of impact for small volumes and an extra concavity (likely due to a selection bias) for very large volumes.  Below we will present two possible explanations for the linear behavior of the impact for small $\phi$. 
\item By considering ${\cal I}(Q,T)$ as a function of both variables, \citep{Zarinelli} introduces the {\it market impact surface} and showed that a double logarithmic function outperforms the power law form of Eq. \ref{eq:fitimpact}.
\end{itemize}

Interestingly, Eq. \ref{eq:fitimpact} can be predicted from the execution of a metaorder with constant participation rate in the continuous time TIM model with $f(v)=\sign(v) |v|^\delta$ and $G(t)=t^{-\gamma}$ with $\delta_T=1-\gamma$ and $\delta_\eta=\delta$, thus $\delta=\gamma=1/2$ \citep{Gatheral2010}.

Notice that the square root impact law is not related to the fact that volatility scales as the square root of (execution) time, which, for a fixed participation rate, is proportional to metaorder size. First, according to definition \ref{eq:impdef}, market impact is a drift term and the inclusion of the metaorder sign $\epsilon$ is critical in the definition, while neglecting $\epsilon$ simply highlights the relation between volatility and volume. 
Second, the result of the regression of Eq. \ref{eq:fitimpact} indicates that, by controlling for both $T$ and $\eta$, market impact is mainly dependent on $\sqrt{T\eta}=\sqrt{\phi}$. Third, as shown explicitly in \citep{BucciJP}, market impact curves of metaorders with $\phi \gtrsim 5\cdot 10^{-4}$ (roughly 80\% of those in the ANcerno database) are independent on $T$ and consistent with a square root dependence on $\phi$. Once more, impact of the remaining small metaorders are better described by a linear relation. \citep{BucciJP} also shows that the variance of impact depends linearly on $T$, as expected by the diffusivity of price, and this price uncertainty largely exceeds the average reaction impact contribution (which in turn explains why the $R^2$ in the market impact estimation is typically very small).

From a modeling perspective, the square root impact law and its deviations are well described by the Locally Linear Order Book (LLOB) model for the coarse-grained dynamics of latent liquidity \citep{DonierLLOB}. In a nutshell, LLOB is a limit order book model whose quantity of interest is the density $\varphi(x,t)$ of latent orders around price $x$ at time $t$. Conventionally, one can choose $\varphi$ to be positive for buy latent\footnote{The LLOB model has been originally developed for describing the latent liquidity, not necessarily the visible one. However close to the spread the two liquidities should coincide.} orders (corresponding to $x < p(t)$, where $p(t)$ is here the current transaction price) and negative for sell latent orders (corresponding to $x > p(t)$). The coarse-grained dynamics of the latent liquidity is well described by
\begin{equation}
\partial_t \varphi= D\partial_{xx} \varphi - \nu \varphi + \lambda~\sign(y)+m~\delta(y) \ ,
\label{eqB}
\end{equation}
where $y\equiv p(t)-x$, and $\nu$ describes order cancellation, $\lambda$ new order deposition and $D\partial_{xx}$ limit price reassessments. The final ``source'' term corresponds to a metaorder of size $Q$ executed at a constant rate $m=Q/T$, corresponding to a flux of orders localized at the transaction price $p(t)$. In the absence of a metaorder ($m=0$), Eq. (\ref{eqB}) admits a stationary solution in the price reference frame, which is linear when $y$ is small, i.e. $\varphi_{\text st}(y) = {\cal L} y$ 
where ${\cal L}=\lambda/\sqrt{D \nu}$ is a measure of liquidity. The total transaction rate $J$ is simply given by the flux of orders through the origin, i.e. $J \equiv D \partial_y \varphi_{\text st}|_{y=0} = D {\cal L}$. 

In the limit of a slow latent order book (i.e. $\nu T \ll 1$), the price trajectory $p_m(t)$ during the execution of the metaorder (obtained as the solution of $\varphi(p_m,t)=0$) is given by the self-consistent expression \citep{DonierLLOB}
\begin{align}
\hspace{2.5cm} p_m(t) =& p_0(t) + y(t), \\  & \hspace{-4cm}y(t) = \frac{m}{{\cal L}} \int_0^t \frac{{\rm d}s}{\sqrt{4 \pi {D} (t-s)}} \exp\left[{-\frac{(y(t) - y(s))^2}{4{D}(t-s)}}\right],
\end{align}
where $p_0(t)$ is the price trajectory in the absence of the metaorder that starts at $t=0$ and ends at $t=T$. 
Interestingly, when impact is small, i.e. if $\forall t,s$ it is $|y(t)-y(s)| \ll D(t-s)$, the above expression for the price dynamics coincides with the TIM with $\delta=1$ and $\gamma=1/2$.

Price impact of a metaorder in the LLOB model is then defined as ${\cal I}(Q,T)=y(T)$, and is found to be given by
\begin{equation}\label{eq:impact}
{\cal I}(Q,T)= \sqrt{\frac{DQ}{J}}\, {\cal F}(\eta) \ ,\quad\mbox{with} \quad \eta\equiv\frac{Q}{JT}\ ,
\end{equation}
where $\eta$ is the participation rate and the scaling function ${\cal F}(\eta) \approx \sqrt{\eta/\pi}$ for $\eta \ll 1$ and $\approx \sqrt{2}$ for $\eta \gg 1$. Hence, ${\cal I}(Q,T)$ is linear in $Q$ for small $Q$ at fixed $T$, and crosses over to a square-root for large $Q$. Note that in the square-root regime, impact is predicted to be independent of the execution time $T$, as approximately observed empirically (see the discussion above). 


 
 The theoretical predictions of LLOB model have been empirically tested in \citep{Bucci2} where, using a large dataset of more than 8 million metaorders from the ANcerno database, it has been shown a remarkable qualitative agreement between the data and the model. However the original model in \citep{DonierLLOB} predicts the crossover of impact from the linear to the square root regime at $\eta^*=1$, while empirical data shows that this value is much closer to $10^{-3}$. \citep{Benzaquen} generalizes the model of \citep{DonierLLOB}  by introducing  (at least) two types of liquidity providers, acting on two different time scales: slow and persistent agents are able to resist the impact of the metaorder and fast agents who lubricate the high-frequency activity of markets. The introduction of two types of agents modifies the value of the crossover participation rate $\eta^*$. \citep{Bucci2} shows that the LLOB model with two types of agents fits quantitatively extremely well the shape of ${\cal I}(Q,T)$ as a function of $\phi$ and $\eta$ when tested on the ANcerno database.

Beside the total impact of a metaorder, it is interesting to investigate the properties of the average price dynamics during and after the execution of a metaorder, because this analysis gives insightful information on the price impact dynamics and the role of information in trading. The first problem has been investigated in  \citep{Zarinelli} by computing the average price path during metaorders' execution by considering subsets of metaorders with different duration $T$ and participation rate $\eta$. Again, large samples are required due to the high level of noise in this type of data and ANcerno dataset allows to perform robust statistical analyses. 
 One of the investigated question is whether, given two metaorders with the same participation
rate $\eta$ and different durations $T_1$ and $T_2$ ($T_1 < T_2$), the market impact reached at time $T_1$ is the same for the two metaorders. The empirical answer is clearly negative: The market impact trajectories deviate from the market impact surface. For small participation rates, this effect is stronger and price trajectories are well above the immediate impact. Moreover, in most cases the price reverts before the end of the metaorder (see also \citep{iuga}), while for larger $\eta$, the price trajectories become closer and closer to the values of the impact surface.
The observation of non-overlapping trajectories might be explained in terms of executions with variable participation rate. A front-loaded execution, i.e., an execution with a decreasing participation rate, produces a strong impact at the beginning and a milder impact toward the end, as observed in real data. This choice might be due to risk aversion \citep{AC} or to the attempt to catch as much liquidity on the book as possible. It is quite interesting to observe that the TIM model with a front-loaded execution is able to reproduce the observed fact that price impact trajectories revert during the execution of the metaorder. On the contrary, a model with permanent impact, such as the Almgren-Chriss model, \citep{AC} always gives monotonic price trajectories if the sign of the trades is uniform. 

The behavior of price after the end of the metaorder is more complicated to estimate, in part because the noise level is even larger than during the metaorder execution. The observed average price dynamics is consistent with a reversion of the price with respect to the value reached at the end of the execution. This is another confirmation of the transient nature of market impact as described, for example, by the TIM. The long term value of the price is even more complicated to estimate for several reasons. First, the very slow decay of impact requires to measure impact on a long time horizon, when volatility effects become dominant. Second, end of day effect and overnight returns could make difficult to estimate permanent impact if the decay continues the days after the metaorder execution. Third, metaorders are sometimes split over multiple days creating an autocorrelation of metaorders, which makes hard to estimate the `bare' decay of price impact. 

From a theoretical point of view, in the `fair pricing' theory of \citep{FGLW} an equilibrium condition is derived between liquidity providers and a broker aggregating informed metaorders from several funds. The theory predicts that the average price payed during the execution is equal to the price at the end of the reversion phase. If metaorder size distribution is power law with tail exponent $3/2$ (as empirically observed), the impact is predicted to decay towards a plateau value whose height is $2/3$ of the peak impact, i.e. the impact reached exactly when the metaorder execution is completed.

Interestingly, several empirical studies reports results compatible with the $2/3$ factor  \citep{moro2009market,Zarinelli,bershova2013non,waelbroeck2013market} although the latter paper notes that the impact of uninformed trades appears to relax to zero. On the other hand, \citep{Brokmann}  underlines the importance of metaorders split over many successive days, as this may strongly bias upwards the apparent plateau value. After accounting for metaorder autocorrelations (from a single fund), the paper concludes that impact decays as a power-law over several days, with no clear asymptotic value. A more extensive analysis has been performed using the ANcerno database in \citep{Bucci3} which shows that while at the end of the same day the average price is on average close to $2/3$ of the peak impact, the decay continues the next days, following a power-law function at short time scales, and apparently converges to a non-zero asymptotic value at long time scales (roughly 50 days) close to $1/3$ of the peak impact.
For such long time lags, however, market noise becomes dominant and makes it difficult to conclude on the asymptotic value of impact, which is a proxy for the (long time) information content of the trades.
 
 \section{Co-impact}\label{sec:coimpact}

In the previous section, market impact of a metaorder is defined by conditioning only on its properties (size and duration). However, in a given day there is typically a large number of funds simultaneously trading the same stock. As empirically observed in \citep{Zarinelli} by investigating the ANcerno database, there is a clear tendency of traders to send metaorders with the same sign (buy or sell) on the same asset. The reason for this coherent behavior are manyfold, but probably the most important one is related to the similarity of trading strategies among institutional investors. One can thus ask how the presence of other metaorders, modifies market impact and the associated  transaction cost of a given metaorder.  This crowding effect on market impact has been termed {\it co-impact} \citep{coimpact}. We are thus changing the conditioning variables in the definition of market impact by considering a vector of simultaneously present metaorders. We will then averaging this quantity by using their joint distribution, keeping as conditioning variable the metaorder whose impact we are interested in. 

\citep{coimpact} investigates how the expected open to close daily logreturn $\Delta p^{(d)}\equiv \log p_{close}/p_{open}$ depends on the order flow generated by the ANcerno metaorders. Consider a day when $N$ metaorders are simultaneously present, each described by $\tilde \phi_i\equiv\epsilon_iQ_i/V$, ($i =1,...,N$), where $V$ is again the average daily volume and $\epsilon_i$ and $Q_i$ are, respectively, the sign and the size of the $i$-th metaorder. Defining the vector $\tilde {\bm \varphi}_N=(\tilde \phi_1,..,\tilde \phi_N)$, the quantity of interest is  
\begin{equation}
I( \tilde{\bm \varphi}_N)\equiv{\mathbb E}[\Delta p^{(d)}| \tilde {\bm \varphi}_N]
\end{equation}
This is however a function of $N$ variables and some parametric restriction must be made to estimate it from data.   \citep{coimpact} empirically finds that the above quantity is well described by $I(\tilde {\bm \varphi}_N)= Y \cdot f_\delta(\Phi)$,
where $\tilde \Phi=\sum_{i=1}^N \tilde \phi_i$ and $f_\delta(v)=\sign(v) |v|^\delta$ with $\delta\simeq1/2$. Thus the average price mainly reacts with a square root law to the total net order flow of ongoing metaorders. This means that the market, due also to the fact that trading is anonymous,  is unable to individually distinguish them. 
 Despite the insensitivity of the price to individual metaorders is quite intuitive, it also raises some issues on how the square root impact can hold.  Let consider a simple example where there is a buy metaorder with order flow $\tilde \phi>0$, which is traded simultaneously with other metaorders with total order flow  $\tilde \phi_m>0$.  Assuming that the square root law applies for the total order flow, the observed impact is
\begin{equation}
{\mathbb E}[\Delta p^{(d)}| \tilde \phi, \tilde \phi_m] \propto  \sqrt{\tilde \phi+\tilde \phi_m}
\end{equation}
Keeping $\tilde \phi_m$ fixed, when  $\tilde \phi \to 0$ market impact tends to a constant, for $\tilde \phi \ll \tilde \phi_m$, instead,  ${\mathbb E}[\Delta p| \tilde \phi, \tilde \phi_m]$ is linear in $\tilde \phi$, and only when $ \tilde \phi \gg \tilde \phi_m$ a square root behavior is expected. Thus, how can a non-linear impact law survive in the presence of a large number of simultaneously executed metaorders?

The argument can be made mathematically more precise by asking what is the expected impact of a metaorder, labeled with $k$, when other $N-1$ are simultaneously being executed. Given the evidence above, this impact can be written as\footnote{Note that we are conditioning on the signed volume fraction $\tilde \phi_k$, which, under buy-sell symmetry, is equivalent to compute the expectation of $\epsilon_k \Delta p^{(d)}$ conditional to absolute volume fraction $\phi$. In other words, also here we are measuring a drift term.}
\begin{equation}
{\mathcal I}_N(\tilde \phi)\equiv{\mathbb E}[\Delta p^{(d)} |\tilde \phi_k=\tilde \phi, N]=Y\int d\tilde \phi_1...d\tilde \phi_N P(\tilde{\bm \varphi}_N |\tilde \phi_k=\tilde \phi) f_\delta(\tilde \phi_k+\sum_{i\ne k} \tilde \phi_i )
\end{equation}
One can then obtain the unconditional impact by averaging ${\mathcal I}_N(\tilde \phi)$  over the distribution $P(N)$ of the number of metaorders per day. Thus  ${\mathcal I}_N(\tilde \phi)$ depends on the joint distribution of order flows $P(\tilde{\bm \varphi}_N)$ and \citep{coimpact} derives the analytical expression for  ${\mathcal I}_N(\tilde \phi)$ under different specification for it (for example multivariate Gaussian). A crossover from a linear to a square root behavior is predicted and the transition point depends on the number of metaorders $N$ and on their correlation (more generally, statistical dependence). When $N$ is small, a small investor will observe linear impact with a non-zero intercept ${\mathcal I_0}$, crossing over to a square-root law at larger $\tilde \phi$. The intercept ${\mathcal I_0}$ grows with the correlation between the signs of the metaorders and can be interpreted as the average impact of all the other metaorders. When the number of metaorders is large and the investor has no correlation with their average sign, one should expect on a given day a square-root impact randomly shifted upwards or downwards by ${\mathcal I_0}$. Averaged over all days, a pure square-root law emerges, which explains why such behavior has been reported in many empirical papers.

Calibrating such model on real data requires to make some assumptions on the joint distribution $P(\tilde{\bm \varphi}_N)$.  \citep{coimpact} shows that the correlation of absolute volume fractions $\phi_i=|\tilde \phi_i|$ is negligible, while correlation between metaorder signs plays an important role. By calibrating a simple heuristic model where a single factor drives the metaorder signs, \citep{coimpact}  reproduces to a good level of precision the different regimes of the empirical market impact curves as a function of $\tilde \phi$, $N$, and the correlation of their signs. In particular, for a metaorder uncorrelated with the rest of the market, the impacts of other metaorders cancel out on average. Conversely, any intercept of the impact law can be interpreted as a non-zero correlation with the rest of the market. 
 
It is interesting to make a comparison with what simple models of market impact predict on price impact when many informed agents are simultaneously present. \citep{Bagnoli} investigates the equilibrium in a one-period Kyle model \citep{kyle1985continuous}.  $N$ symmetrically and informed agents trade one asset in a market where uninformed agents and market makers are also present.  \cite{Bagnoli} shows that the Kyle's lambda, i.e. the proportionality factor between price impact and aggregated order flow,  scales as $N^{-1/\alpha}$, where $\alpha$ is the exponent of the stable law describing the price and uninformed order flow distribution. Moreover if the second moment of both variables is finite, \cite{Bagnoli} shows that the Kyle's lambda scales as $1/\sqrt{N}$. Interestingly,  Figure 3 of \cite{coimpact}  shows that market impact of a ANcerno metaorder decreases with the number of metaorders simultaneously present.

From a practical perspective, the model and the empirical observations are important for traders to estimate (pre- and post-execution) the cost of their trades, and thus to help them deciding when is the right moment to trade. For example, \citep{Lehalle}, investigating the ANcerno database, finds an approximately linear relation between the implementation shortfall of a metaorder and the net trading imbalance due to the other metaorders simultaneously traded. When the trade is in the same direction as the net order flow imbalance,  one could expect  to pay a significant trading cost up to $0.4$ points of price volatility, while one could expect to benefit from a price improvement of $0.3$ points of volatility when the trader is almost alone in front of his competitors aggregate flow. In a normal trading situation, the information on the ongoing metaorders is not available, thus statistical and machine learning methods could be used to infer, at least partly, this information from the visible order flow.

\section{Conclusion}\label{sec:conclusions}

As it should be clear from this short review, in the last twenty years we have made huge progresses in understanding the important and fascinating problem of how price is formed in financial markets as the result of order flow and trading activity. This advancement is due to the availability of very detailed and rich datasets and to the development of sophisticated models able to capture, at least partly, the strong dependencies and feedbacks between orders and prices. Still much remains to do. For example, most models are inherently stationary and with fixed parameters, while liquidity, as many market variables, are highly dynamic and latent. Methods from econometrics (filtering, score driven models) and machine learning (reinforcement learning) can provide the tools for tackling this important aspect of market dynamics. Combining these models with optimal execution or optimal market making solutions available in real time would certainly provide a great addition for the industry.

\bibliographystyle{apalike}
\bibliography{subchapter_Lillo.bib} 

\end{document}